\documentstyle[aps,multicol,epsfig]{revtex}

\def\br{ {\bf r} }

\def\bk{ {\bf k} }
\def\bv{ {\bf v} }
\def\bps{ {\bf p}_s }

\def\im{ \,{\rm Im}\, }
\def\re{ \,{\rm Re}\, }

\begin{document}
\bibliographystyle{unsrt}
\draft

\title{Effect of magnetic field on impurity bound states in high-$T_c$
superconductors}

\author{K.~V.~Samokhin$^*$ and M.~B.~Walker}

\address{Department of Physics, University of Toronto, Toronto, Ontario,
 Canada M5S 1A7}

\date{\today}

\maketitle

\begin{abstract}
We consider the influence of a magnetic field $H$ on the quasiparticle bound
states near scalar impurities in $d$-wave superconductors. A ``Doppler shift''
in the excitation energies induced by the supercurrent leads to several
important effects. At large but finite impurity strength,
there are corrections to the energy and width of the impurity-induced resonance,
proportional to $H^2$. On the other hand, in the limit of very strong
impurity potential (unitary limit), the  bound state is destroyed
and acquires a finite width proportional to $H(\ln H)^{-1}$. There are also
considerable changes in the asymptotic behaviour of the bound state wave functions.
\end{abstract}


\begin{multicols}{2}\narrowtext

High-temperature superconductors
(HTSC) belong to a class of unconventional superconductors with
$d$-wave symmetry \cite{dwave}. A non-trivial orbital structure
of the order
parameter, in particular the presence of the gap nodes, leads to the
effects of disorder in HTSC being much richer than in conventional
materials. For instance, in contrast to the $s$-wave case, the
Anderson theorem does not work and non-magnetic impurities exhibit
a strong pair-breaking effect. Also, a finite concentration of
disorder produces a non-zero density of quasiparticle states at zero
energy, which results in a considerable
modification of the thermodynamic and transport properties at low
temperatures \cite{Book}.

One of the striking features of $d$-wave superconductors is that
even a single scalar impurity has a notable effect on the
superconducting state, creating a quasiparticle impurity bound state (IBS)
in its vicinity whose energy and width tend to zero in the limit of strong
impurity potential \cite{Balat95}. Various aspects of the theory
of IBS have been elaborated by many authors, both analytically
\cite{Balat95,Salk96,SBS97} and numerically \cite{IBS-numer}.
From the practical point of view, the high sensitivity of IBS to the symmetry
of the order parameter and the impurity
strength makes these states a powerful tool for probing the
properties of HTSC. This has been done
recently in beautiful scanning tunneling microscopy (STM)
experiments on BSCCO compounds \cite{Pan00,Huds99,Yazd99}, in
which many of the theoretically predicted  features of IBS were observed,
such as (i) sharp peaks in the energy dependence of the
single-particle density of states (DoS), (ii) a cross-shape anisotropy of
the quasiparticle wave functions in $a$-$b$ plane, and (iii) a slow
power-law decay of the wave functions away from the impurity.
(Some apparent discrepancies
between the theory and the experiment, such as the
position of the maximum local DoS directly above the impurity site
as seen in STM pictures, can be attributed to the blocking effect
of Bi-O layers \cite{ZT00} and thus do not invalidate the
standard theoretical picture of IBS.)

All previous studies have neglected the influence
of an external magnetic field on IBS. It is known, however, that
in the presence of the gap nodes, the supercurrent induced by a magnetic field
acts as an effective pair breaker and
gives rise to a finite density of bulk quasiparticles at low energies. This should
affect the properties of IBS because of the hybridization of localized and delocalized
states. The purpose of the present article is to study the combined effect
of magnetic field and strong impurity scattering in $d$-wave superconductors.
We neglect the Zeeman splitting and concentrate
on the physical properties relevant to the $c$-axis STM experiments which
probe the local DoS at various distances from an impurity site.
The influence of the magnetic field on the disorder-averaged density of states
in $d$-wave superconductors with randomly distributed unitary impurities was
studied in Ref. \cite{Barash97}.

Let us consider a repulsive point-like impurity with potential
$U(\br)=u\delta(\br)$ ($u>0$) in a two-dimensional
$d$-wave superconductor. The external magnetic field can be directed either
in the $a$-$b$ plane or along the $c$ axis. In both cases, we
assume that the supercurrent $\bps$ is locally uniform,
and use the gauge in which the order parameter is real \cite{real}.
The model Hamiltonian can be written in the following form:
\begin{equation}
{\cal H}=\sum\limits_{\bk\bk'} C^\dagger_\bk
 H_{\bk\bk'} C_{\bk'},
\end{equation}
where $C_\bk=(c_{\bk\uparrow},c^\dagger_{-\bk\downarrow})^T$
are Nambu operators, and
\begin{equation}
\label{Hbdg}
H_{\bk\bk'}=\left(\begin{array}{cc}
H_+ & \Delta_\bk \\
\Delta_\bk & -H_-
\end{array}\right).
\end{equation}
is the Bogoliubov-de Gennes (BdG) operator. The normal part
of $H_{\bk\bk'}$ depends on the supercurrent via the gauge
transformation:
$H_\pm=\xi_{\bk\pm\bps}+U_{\bk-\bk'}\simeq\xi_\bk\pm{\bf v}_F\bps
+U_{\bk-\bk'}$, where $\xi_\bk=k^2/2m-\mu$ is the normal state
spectrum (we assume a spherical Fermi surface, which does not restrict
the generality of our results, but considerably simplifies the
calculations), ${\bf v}_F=\nabla_\bk\xi_\bk$ is the Fermi velocity, and
$U_\bk$ is the Fourier transform of the impurity potential.
The mean-field order parameter,
corresponding to $d_{x^2-y^2}$ symmetry, has the form
$\Delta_\bk=\Delta_0(\hat k_x^2-\hat k_y^2)=\Delta_0\cos 2\varphi$,
where $\varphi$ is the azimuthal angle in $a$-$b$ plane.  Note
that we do not calculate the order parameter self-consistently and
assume $\Delta_0$ to be constant. The numerical
investigation of the self-consistent BdG equations shows that there
is some suppression of the order parameter magnitude
near an impurity site \cite{Hett99}, which can affect the low-energy
behaviour of DoS in a system with a finite number of
impurities \cite{Atkins00}.
Here we study the single-impurity limit, which corresponds to a dilute
disorder concentration. In this case, it has been demonstrated
in Ref. \cite{Shnir99} that the order parameter variation leads only
to renormalization of the effective impurity strength towards
the unitary limit.

The quantity measured in STM experiments is the local differential
conductance $dI/dV$, which is proportional to the local density of
states:
\begin{equation}
\label{DoSdef}
 N(\br,\omega)=-\frac{1}{\pi}\im G^R_{11}(\br,\br;\omega),
\end{equation}
where $G^R$ is the retarded Gor'kov-Nambu matrix Green's function.
In the presence of a single scalar impurity, one can express
$G^R$ in terms of the Green's function of a clean superconductor.
Thus,
\begin{eqnarray}
\label{G}
 G^R(\br_1,\br_2;\omega)&=&G^R_0(\br_1-\br_2,\omega)
 \nonumber\\
 &&+G^R_0(\br_1,\omega)T(\omega)G^R_0(-\br_2,\omega),
\end{eqnarray}
where the $T$-matrix is given by
$T(\omega)=u\tau_3\left[1-ug(\omega)\tau_3\right]^{-1}$, and
$g(\omega)=G^R_0({\bf 0},\omega)=\int d^2k/(2\pi)^2\;G^R_0(\bk,\omega)$.
The Green's function $G_0^R$ describes a homogeneous system without
an impurity, but in the presence of a uniform supercurrent. The effect of
the supercurrent in a translationally invariant system amounts to
a ``Doppler shift'' in the quasiparticle energy. In the
momentum representation, we have
\begin{equation}
\label{G_0}
 G^R_0(\bk,\omega)=\frac{(\omega_+-\bv_F\bps)\tau_0+\xi_\bk\tau_3+
 \Delta_\bk\tau_1}{(\omega_+-\bv_F\bps)^2-\xi^2_\bk-\Delta^2_\bk},
\end{equation}
where $\omega_+=\omega+i0$, and $\tau_i$ are the Pauli matrices.

The energies of the impurity-induced IBS correspond to the poles of
the $T$-matrix and satisfy the equation $\det T(\omega)=0$, whose
solutions can be complex. We are
interested in the limit of small supercurrent and strong impurity
scattering, so that the relevant energies are expected to be small
compared to the gap magnitude. This case is of
particular interest because the most profound effects related to
IBS have been observed in the vicinity of Zn impurities in
BSCCO, which have the $s$-wave phase shift $\delta_0\simeq
0.48\pi$ and thus are very close to the unitary limit
\cite{Pan00}.
The momentum integrals can be easily calculated,
giving the following result at $\im\omega=0$:
\begin{equation}
\label{im_g}
\im g(\omega,\bps)=g_0(\omega,\bps)\tau_0+g_1(\omega,\bps)\tau_1,
\end{equation}
where, in leading order in $(v_Fp_s/\Delta_0)^2$,
\begin{eqnarray}
&& \label{g_0} g_0=-\frac{\pi N_F}{4\Delta_0}
   \sum\limits_{i=1}^4|\omega-{\bf v}_i\bps|,\\
&& g_1=-\frac{\pi N_F}{8\Delta_0^2}
 [\hat{\bf z}\times\bps]\cdot\sum\limits_{i=1}^4
 (-1)^i\bv_i|\omega-\bv_i\bps|. \nonumber
\end{eqnarray}
In these expressions, $N_F$ is the DoS in the normal state at the
Fermi level \cite{ehasymm}, $i$ labels the gap nodes, and ${\bf v}_i$ are the
Fermi velocities at the nodes. We see that, at small supercurrents,
it is possible to neglect the off-diagonal terms in (\ref{im_g})
(retaining these terms would give the corrections of the order
of $(v_Fp_s/\Delta_0)^4$ to the results below).
From the Kramers-Kronig relations, we obtain
\begin{equation}
\label{re_g}
 \re g(\omega)=-\frac{N_F}{2\Delta_0}
  \sum\limits_{i=1}^4\;(\omega-{\bf v}_i\bps)
   \ln\frac{\Delta_0}{|\omega-{\bf v}_i\bps|}\;\tau_0.
\end{equation}

The next step in the derivation of the $T$-matrix is to continue
$g(\omega)$, whose real and
imaginary parts at the real axis are given by Eqs. (\ref{re_g}) and
(\ref{im_g}), to the whole complex plane of $\omega$. Introducing
the notation $z=\omega/\Delta_0$, we have $g(\omega)=\pi N_FF(z)\tau_0$, where
\begin{equation}
\label{F_def}
F(z)=\frac{1}{2\pi}\sum\limits_{i=1}^4\;(z-z_i)\ln(z-z_i)-iz
\end{equation}
with $z_i={\bf v}_i\bps/\Delta_0$ ($\sum_i z_i=0$). Four
logarithmic branch cuts go down from $z=z_i$ parallel
to the negative imaginary axis. Finally,
\begin{equation}
\label{Tmatrix}
T(z)=\left( \begin{array}{cc}
 \displaystyle \frac{u c}{c-F(z)} & 0 \\
  0 & \displaystyle -\frac{u c}{c+F(z)}
\end{array} \right),
\end{equation}
where $c=(\pi uN_F)^{-1}=\cot\delta_0>0$ for a repulsive impurity.
This expression for the $T$-matrix is valid at $|z|,|z_i|\ll 1$.

The IBS spectrum is determined by the equation
\begin{equation}
\label{eqF}
 F(z)=\pm c.
\end{equation}
In the absence of a supercurrent, $F(z)\to F_0(z)=(2/\pi)z\ln z
-iz$, and Eq. (\ref{eqF}) can be easily solved at $c\ll 1$, giving
the IBS energy of the form $\omega_0=z_0\Delta_0$, where
$\re z_0=\mp \pi c/2|\ln c|$, $\im z_0=-\pi^2c/4\ln^2c$
with logarithmic accuracy \cite{Balat95}.
The presence of a non-zero imaginary part indicates that the
impurity-induced bound state is in fact a narrow resonance.

At $p_s\neq 0$, the effect of supercurrent on IBS
strongly depends on the relation
between the Doppler shift $v_Fp_s$ and the ``bare'' energy $\omega_0$.
If $\bps$ is parallel to one of the crystallographic axes, e.g. $(100)$,
then $z_{1,4}=-z_{2,3}=z_s=v_Fp_s/\sqrt{2}\Delta_0$, and Eq. (\ref{F_def})
takes the form
\begin{eqnarray}
\label{F1}
 F(z)=-iz+\frac{1}{\pi}(z+z_s)\ln(z+z_s)\nonumber\\
 +\frac{1}{\pi}(z-z_s)\ln(z-z_s).
\end{eqnarray}
For $z_s\ll|z_0|$, the solution
of Eq. (\ref{eqF}) can be found perturbatively in $z_s$.
Using the expansion $F(z)=F_0(z)+z_s^2/\pi z$, we find
\begin{equation}
\label{p_ssmall}
\left\{\begin{array}{l}
 \displaystyle \re z=\re z_0\mp\frac{1}{\pi c} z_s^2,\\
 \displaystyle \im z=\im z_0-\frac{1}{2c\ln^2c} z_s^2.
\end{array}\right.
\end{equation}
These expressions are valid as long as $|\delta z(z_s)/z_0|\ll 1$.
Thus, in the presence of a small
magnetic field, the corrections to the IBS energy and width are
proportional to $H^2$ ($z_s^2=0.5[H/H_c(0)]^2$ for ${\bf H}\perp\hat z$).

More interesting is the opposite limit of ``large''
supercurrents, which is relevant when the impurity scattering is
close to the unitary limit.
In zero field, the bare IBS energy vanishes at $c=0$, so the impurity-induced
resonance gets decoupled
from the continuum of propagating excitations and becomes a true
bound state with zero width. The dominant energy scale in this
case is provided by the Doppler shift $v_Fp_s$, which makes it possible
to treat the right-hand side of Eq. (\ref{eqF}) as a small perturbation.
It can be checked that the equation $F(z)=0$ has only one solution in the
complex plane:
\begin{equation}
z_*=-i\frac{\pi z_s}{2|\ln z_s|}.
\end{equation}
At $c\neq 0$, we look for a solution in the form $z=z_*+\delta
z(c)$ and obtain
\begin{equation}
\label{p_slarge}
\left\{ \begin{array}{l}
 \displaystyle \re z=\mp\frac{\pi c}{2|\ln z_s|}, \\
 \displaystyle \im z=-\frac{\pi z_s}{2|\ln z_s|}+O(c^2).
\end{array}\right.
\end{equation}
These expressions are valid as long as $|\re z/z_*|\ll 1$, i.e. at $z_s\gg c$.
From Eqs. (\ref{p_slarge}) we see that the zero-energy IBS in the
unitary limit is destroyed by magnetic field, getting replaced by a resonance
whose width is proportional to $H(\ln H)^{-1}$.
The physical reason for this is clear from Eqs. (\ref{DoSdef}) and (\ref{im_g}):
at non-zero supercurrent, the DoS of the bulk excitations does not
vanish at $\omega=0$, which leads to a stronger hybridization of the IBS
and the continuum of propagating states. We would like to stress that the result
(\ref{p_slarge}) is manifestly non-perturbative in magnetic field.

The calculations for other directions of supercurrent can be done in
a similar fashion.
Solution of Eq. (\ref{eqF}) leads to the expressions analogous to
(\ref{p_ssmall}) and (\ref{p_slarge}), albeit
with different numerical coefficients. Therefore, the qualitative effect
of magnetic field on the IBS energy does not depend on the direction of
$\bps$.

\begin{figure}
\begin{center}
\leavevmode
\epsfxsize=0.9\linewidth
\epsfbox{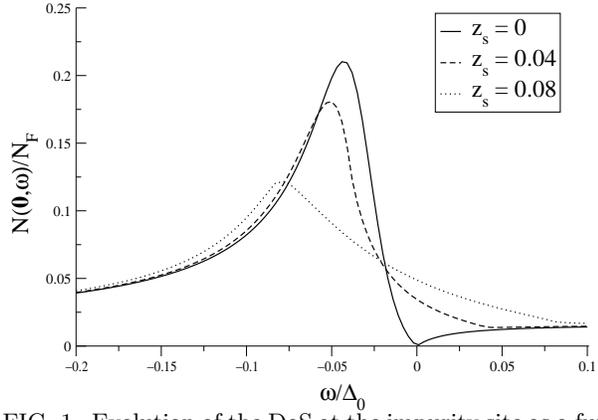}
\caption{Evolution of the DoS at the impurity site as a function
of energy at increasing magnetic field ($\bps\parallel{\bf a},\;c=0.1$).}
\label{Fig1}
\end{center}
\end{figure}

In order to visualize our results and facilitate the comparison with
STM experiments, we have computed the local DoS at the impurity site:
\begin{equation}
\label{N0}
 N({\bf 0},\omega)=-N_F\im\frac{cF(z)}{c-F(z)},
\end{equation}
where $F$ is given by Eq. (\ref{F_def}).
The dependence of $N({\bf 0},\omega)$ on energy and supercurrent
is plotted in Fig. \ref{Fig1}. We see that the asymmetric
peak in the local DoS, which corresponds to a hole-like resonance at
$c>0$, gets shifted and broadened in the presence of supercurrent.

\begin{figure}
\begin{center}
\leavevmode
\epsfxsize=0.9\linewidth
\epsfbox{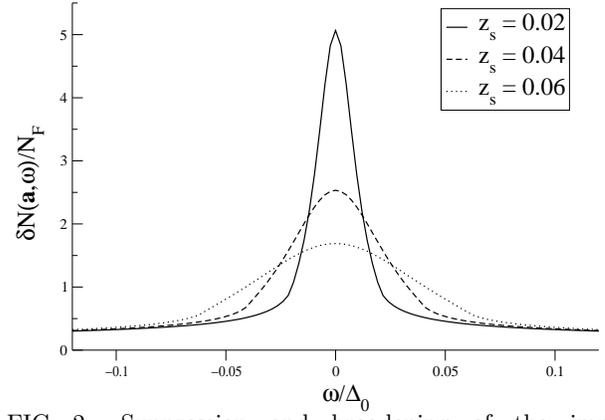}
\caption{Suppression and broadening of the impurity-induced DoS
peak at a nearest neighbor site in the unitary limit at increasing
magnetic field for $\bps\parallel{\bf a}$ (a sharp peak at zero field
is not shown).}
\label{Fig2}
\end{center}
\end{figure}

It follows from (\ref{N0}) that $N({\bf 0},\omega)=0$ at
$c=0$. For this reason, the STM measurements directly at
the impurity site are not useful in the unitary limit. To study the
effect of supercurrent on IBS in this case, one should calculate
the local DoS away from the point $\br=0$, e.g. at one of the nearest
neighbors of the impurity site, where the local DoS is known to reach
its maximum value at $p_s=0$ \cite{Salk96}.
According to Eqs. (\ref{DoSdef}), (\ref{G}) and (\ref{Tmatrix}),
the local DoS at site $\br$ in the unitary limit can
be represented in the form
\begin{eqnarray}
\label{Nr}
 N({\bf r},\omega)&=&N_0(\omega)\nonumber\\
 &&+\frac{1}{\pi^2N_F}\im\frac{1}{F(z)}
 \left[G_0^R(\br,\omega)G_0^R(-\br,\omega)\right]_{11},
\end{eqnarray}
where $N_0(\omega)=-(1/\pi)g_0(\omega)$ is the DoS for a clean
$d$-wave superconductor in the presence of supercurrent, and
$G_0^R(\br,\omega)$ is the Fourier transform of Eq. (\ref{G_0}).
At $\br={\bf a}$ and $(\omega,v_Fp_s)/\Delta_0\to 0$, one
can neglect the first term on the right hand side of Eq. (\ref{Nr})
compared to the second one,
which is singular in this limit. The singularity comes from
$F^{-1}(z)$, while the product of two Green's functions is
not singular and can be replaced by its value at $\omega=p_s=0$,
which is real. Therefore, $N({\bf a},\omega)/N_F\sim
\im F^{-1}(z)$.
We have calculated the energy
dependence of $\delta N({\bf a},\omega)=N({\bf a},\omega)-N_0(\omega)$
for $\bps\parallel{\bf a}$ and
plotted the results in Fig. \ref{Fig2}. As the supercurrent increases,
so does the width of the zero-energy peak, whereas its magnitude decreases.
Experimentally, a notable suppression of the $c$-axis zero-bias conductance
peaks (ZBCP) has been reported in Ref. \cite{Ekin97} for YBCO/Ag and
TBCCO/Au planar junctions. It is tempting to attribute this observation to
the effect of magnetic field on the IBS induced by strong defects
at the surface of those materials.

Another peculiar property of IBS, which was
predicted in Ref. \cite{Balat95} and observed in recent
STM experiments \cite{Pan00,Huds99}, is a sharp four-fold anisotropy
of the IBS wave functions at zero bias, with a characteristic $1/r^2$-tails
at large distances from the impurity. This not only reflects the microscopic symmetry
of superconducting state, but also has important consequences for the quasiparticle
transport in HTSC.
It was argued in Ref. \cite{BalSal96} that, in the presence
of many impurities, the overlap of the extended IBS wave
functions along the gap node directions can result in the formation of
an impurity band, where all states are delocalized and can thus participate
in quasiparticle transport. This should be
contrasted to the propagating excitations which are localized in the
presence of short-range disorder potential \cite{Lee93}.

Here we address the following question: What is the effect of magnetic field
on the asymptotics of the IBS wave functions in the unitary limit? We
study the decay of $\delta N(\br,\omega)=N(\br,\omega)-N_0(\omega)$
along the nodal direction $(110)$  at $\omega=0$.
The coordinate dependence of $G_0^R(\br,\omega)$ in Eq. (\ref{Nr})
is described by rather cumbersome expressions (the details of calculations will
be given in a separate publication), which take on a particularly simple
form in the limit of large distances from the impurity $r\gg p_s^{-1}
\gg \xi_0=v_F/\Delta_0$. In contrast to our results for the IBS energies,
the asymptotic behaviour of the wave functions strongly depends on the
direction of the supercurrent. If $\bps$ is perpendicular to the nodal direction
$(110)$, then, in leading order in $1/r$,
\begin{equation}
\label{dN1}
 \delta N(\br,0)=\frac{\gamma(p_s) N_F\xi_0^2}{4\pi^2}
 \frac{\sin^2 k_Fr}{r^2},
\end{equation}
for $k_F\xi_0\gg p_sr\gg 1$, and
\begin{equation}
\label{dN2}
 \delta N(\br,0)=-\frac{\gamma(p_s) p_s}{8\pi^2\Delta_0}
 \frac{1}{r},
\end{equation}
for $p_sr\gg k_F\xi_0\gg 1$ (we introduced the notation
$\gamma(p_s)=\im F^{-1}(z=0)$). If $\bps$ is parallel to $(110)$, then
\begin{equation}
\label{dN3}
 \delta N(\br,0)=-\frac{N_Fp_s\xi_0^2}{8\pi}\;
 {\rm Im}\left[F^{-1}(0)\frac{e^{2ik_Fr}}{r}\right].
\end{equation}
Eq. (\ref{dN1}) is essentially the
zero-field result, whereas Eqs. (\ref{dN2},\ref{dN3}) show that magnetic field
leads to a slower power-law decay of the envelope IBS wave functions along the nodal
directions, or even the disappearance of the Fermi oscillations at very large distances
from the impurity. These changes reflect the contributions from the nodal
quasiparticles excited by supercurrent. In particular, the non-oscillatory
tail in Eq. (\ref{dN2}) comes from the $(\bar 110)$ and $(1\bar 10)$ nodal states for
which $\bk_F\br=0$.

In conclusion, we have studied the influence of magnetic field on the impurity
bound states in $d$-wave superconductors. We have found several effects, which
can be directly measured in STM experiments, such as a non-linear shift of the
electron-hole asymmetric peak in the local DoS at $c\neq 0$, and also strong suppression
and broadening of the zero-bias peak in the unitary limit. The changes in the
wave function asymptotics can lead to a stronger long-range overlap between
the bound states at different impurities, which might considerably affect the quasiparticle
transport in HTSC.
\\

We are pleased to thank John Wei for useful
discussions. This work was supported by the Natural Sciences and Engineering
Research Council of Canada.

\end{multicols}



\begin{references}
\bibitem{real}
These assumptions are strictly valid only for the in-plane
orientation of ${\bf H}$. However, since we consider
the bound states localized near the impurity,
the inhomogeneity of the supercurrent distribution can be neglected
even in the presence of vortices (as long as we are not close to
$H_{c2}$ and the impurity site is far from a vortex core).
\bibitem{ehasymm}
Here we assume an electron-hole symmetric band.
For an asymmetric band, the situation is more complicated, in particular,
the dependence of the IBS spectrum on the scattering amplitude $u$ is different
(see Refs. \cite{IBS-numer}).
\end{references}
\end{document}